\documentclass[finalsubmission, journal]{IEEEtran}
\usepackage{amsmath,amsfonts}
\usepackage{algorithmic}
\usepackage{algorithm}
\usepackage{array}
\usepackage[export]{adjustbox} 
\usepackage{subcaption}
\usepackage{textcomp}
\usepackage{stfloats}
\usepackage{url}
\usepackage{verbatim}
\usepackage{graphicx}
\usepackage[space]{cite}
\usepackage{color} 
\usepackage{soul}

\begin{document}

\title{The optical architecture of a heterogenous quantum network deployed in production facilities}

\author{Alberto~Sebasti\'an-Lombraña, Hans~H.~Brunner, David~Rinc\'on, Juan~P.~Brito, Rub\'en~B.~M\'endez, Rafael~J.\,Vicente, Jaime~S.~Buruaga, Laura~Ortiz, José~L.~Rosales, Chi-Hang~Fred~Fung, Momtchil~Peev, Jos\'e~M.~Rivas-Moscoso, Felipe~Jim\'enez, Antonio~Pastor, Diego~R.~L\'opez, Jes\'us~Folgueira, C\'esar~S\'anchez and Vicente~Mart\'in
\thanks{Alberto~Sebasti\'an-Lombraña, Juan~P.~Brito and Vicente~Mart\'in are with Universidad Polit\'ecnica de Madrid, Madrid, Spain; with Dept. LSIIS, at the E.T.S.I.Inf, Boadilla del Monte, Spain; and with Center for Computational Simulation, Madrid, Spain.}
\thanks{Rub\'en~B.~M\'endez, Rafael~J.~Vicente, Jaime~S.~Buruaga and José~L.~Rosales are with Universidad Polit\'ecnica de Madrid, Madrid, Spain; and with Center for Computational Simulation, Madrid, Spain.}
\thanks{Laura~Ortiz is with Universidad Polit\'ecnica de Madrid, Madrid, Spain; with Dept. DATSI, at the E.T.S.I.Inf, Boadilla del Monte, Spain; and with Center for Computational Simulation, Madrid, Spain.}
\thanks{Hans~H.~Brunner, Chi-Hang~Fred~Fung, and Momtchil~Peev are with Munich Research Center, Huawei Technologies Duesseldorf GmbH, Munich, Germany.}
\thanks{David~Rinc\'on and C\'esar~S\'anchez are with IMDEA Software Institute, Pozuelo de Alarcón, Spain.}
\thanks{Jos\'e~M.~Rivas-Moscoso, Felipe~Jim\'enez, Antonio~Pastor, Diego~R.~L\'opez and Jes\'us~Folgueira are with Telef\'onica gCTIO/TID, Madrid, Spain.}
\thanks{Manuscript created July, 2025.}}

\markboth{}%
{Sebastian-Lombraña \MakeLowercase{\textit{et al.}}: The optical architecture of a heterogenous quantum network deployed in production facilities}


\maketitle

\begin{abstract}
Quantum Communications promise advances in cryptography, quantum computing and clock synchronisation, among other emerging applications. 
However, communication based on quantum phenomena requires an extreme level of isolation from external disturbances, complicating the co-propagation of quantum and classical signals.
The challenge is greater when deploying networks that are both heterogeneous (e.g., multiple vendors) and installed in production facilities, given that this type of infrastructure already supports networks loaded with their own requirements.
Moreover, to achieve a broad acceptance among network operators, the joint management and operation of quantum and classical resources, compliance with standards, and legal and quality assurance need to be addressed.

This article presents solutions to the aforementioned challenges validated in the Madrid quantum network during the implementation of the projects CiViC and OpenQKD.
This network was designed to integrate quantum communications in the telecommunications ecosystem by installing quantum-key-distribution modules from multiple providers in production nodes of two different operators.
The modules were connected through an optically-switched network with more than 130~km of deployed optical fibre.
The tests were done in compliance with strict service level agreements that protected the legacy traffic of the pre-existing classical network.
The goal was to ensure full quantum-classical interoperability at all levels, while limiting the modifications to optical transport and encryption and complying with relevant standards.
This effort is intended to lay the foundation for large-scale quantum network deployments. 
\end{abstract}


\section{Introduction}
\label{sec:int}

\IEEEPARstart{T}{here} is a growing interest in the use of quantum-enhanced communication services, at the moment mainly based on quantum key distribution (QKD)~\cite{qkd_wiley}.
However, this requires the deployment of a quantum communication infrastructure (QCI) that is compatible with the already deployed infrastructure for classical communication.

This article provides a detailed analysis of the optical infrastructure of the quantum network deployed in Madrid~\cite{madrid_corto}.
This covers the optical design, its implementation, and its evaluation, incorporating key lessons learned from subsequent QCI deployments in Madrid.

The principles of quantum mechanics imply that the measurement of a quantum signal alters it.
Therefore, passive networks, without active elements such as amplifiers, must be used for their transmission.
Even so, the propagation of quantum signals in physical media---typically the optical fibre---entails a probability of absorption that severely limits their range.
If the medium is shared, other quantum or classical signals can further limit the range and, in the case of optical fibres, their deployment is often the most expensive part.
For these reasons, until quantum repeaters are developed~\cite{review_qinternet}, it is necessary for quantum networks to handle these limitations.

Equally relevant to the success of quantum communications is achieving quantum networking that enables the full provision of quantum-enhanced services.
This involves many well-known network functionalities, but one notable feature is the joint operation and management of both quantum and classical resources installed on the network.
Low-level enabling techniques are also necessary to share the underlying physical infrastructure.
To achieve widespread deployment, interoperability of deployed systems, standardisation, certification of quantum networks and devices, compliance with regulations, and quality assurance, are also essential.
\IEEEpubidadjcol

The Madrid quantum network was designed to be heterogeneous at both the quantum and classical levels in order to address these problems comprehensively.
It was also large and complex enough to significantly advance the current state of the art.
To build a diverse network and conduct a comprehensive investigation, multiple technologies from different vendors were installed, including 26~QKD modules from three different manufacturers, some using continuous-variable QKD (CV-QKD) and others discrete-variable QKD (DV-QKD).

\begin{figure}[h]
\centering
\includegraphics[width=\linewidth]{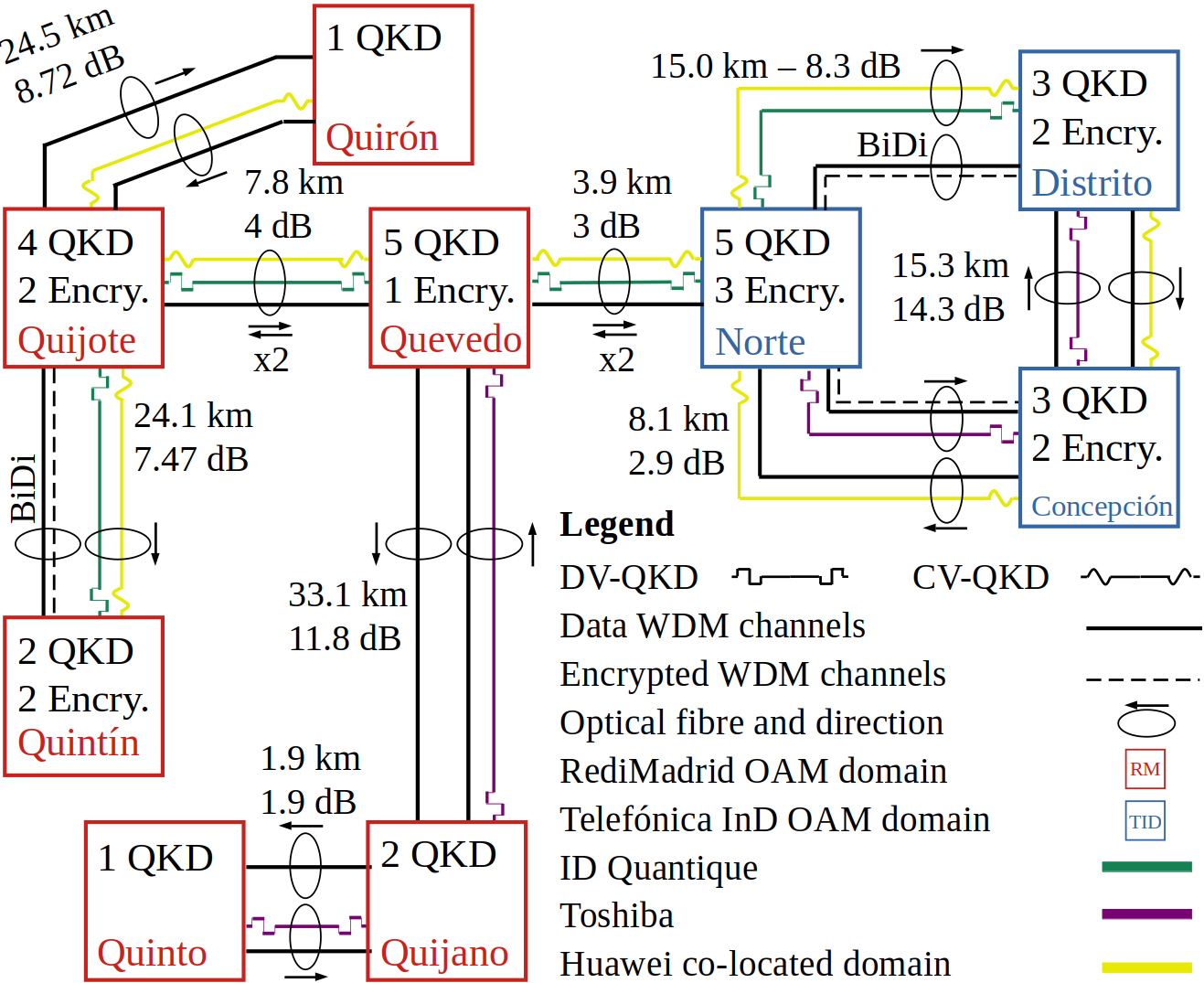}
\caption{Madrid quantum network overview. 26~QKD modules from different manufacturers were installed in the 9~production facilities of RM and TID. More than 130~km of fibre optic pairs supported both classical and quantum signals including, in some cases, third party traffic. Three OSI level-1 encrypted links were deployed, as well as six level-2 Ethernet encrypting devices. A single pair of optical fibres was used in each inter-node connection, but different schemes of coexistence were implemented. The link between Quevedo and Norte was co-managed as a border connection.}
\label{fig:madqci}
\end{figure}

The complexity of the network was addressed by using administrative domains.
The infrastructure was hosted in production premises of two network operators: Telef\'onica Spain, through Telef\'onica Innovaci\'on Digital (TID), the research and development division of the incumbent telecommunications operator in Spain, and the research and education network operator in the region of Madrid, REDIMadrid (RM). 
In addition, a third (logical) domain was used to handle the testing of a switched QKD network.
This specific sub-network is referred to as the ``co-located domain'' and contains all the devices that support switching~\cite{madrid_hwdu_switched, hwdu_cvqkd}.
The network configuration resembles production networks---including an inter-domain border link---reusing already installed resources and adding new ones as required. 
Fig.~\ref{fig:madqci} depicts the layout of that network.

The operation of the global quantum network was based on a standardized key management system combined with a software-defined approach to quantum networks (SD-QKD)~\cite{madrid_corto}, which allowed cryptographic keys to be distributed between any pair of nodes in the network, regardless of the QKD manufacturer and network domain.
Moreover, this ease the accommodation of the heterogeneous QKD systems, which varied in key rates and standardization compliance.
These capabilities were made available to applications through standardised interfaces and the execution of a wide variety of use cases based on quantum-distributed keys.
Specific hardware was installed to provide encryption at OSI layers~1 and~2, as well as software applications to allow secure communication on higher OSI layers, such as IPsec~\cite{madrid_ipsec_rubén} and TLS~\cite{madrid_tls_jaime} implementations.

The paper is organised as follows. 
In the next sections, the related work and an overview about quantum and optical communications is outlined and the coexistence and integration approach of classical and quantum optical networks is discussed. 
Then, a detailed description of the solutions implemented in the Madrid quantum network and its main performance is presented.
The paper finishes with the conclusions.

\section{Related work}
\label{sec:rel-wor}

Quantum cryptography was the first major technological realisation of quantum communication and QKD has grown around the idea of using optical communications and networks since its earliest days~\cite{review_qinternet}.
Examples include the Boston testbed, which already involved optical networking techniques such as switching or dense wavelength division multiplexing (DWDM), as well as other well-known examples such as the Vienna, Geneva and Tokyo quantum networks, which used dark optical fibres deployed in the field.
The first iterations of the Madrid quantum network also date back to that time, along with others such as the Hefei testbed.
This early deployment later evolved into the Hefei–Shanghai–Beijing backbone QKD network, one of the most extensive and operationally ambitious quantum networks to date.

Building on these foundations, many technological improvements over the last decade have brought QKD to a commercial status.
Furthermore, since 2019, a broad effort is underway to establish the European Quantum Communications Infrastructure (EuroQCI), a pan-European quantum network focused on secure communications. 
Currently, several efforts are underway to deploy quantum communication infrastructures (QCI) in Europe. 
A list of these can be found in~\cite{petrus}.

The next big step in the adoption of quantum communication will be the leap to the space segment, since it would enable worldwide quantum connectivity.
Also, further progress is being made in other types of quantum communications such as quantum entanglement distribution---to enable the so-called ``quantum internet''~\cite{review_qinternet, review_stacks}---or other quantum-enhanced applications such as quantum clock synchronisation. 

\section{Quantum and optical communications fundamentals}
\label{sec:com-ove}
This section introduces some topics in telecommunications and quantum technologies, particularly those using optical fibres.

\subsection{Quantum key distribution and cryptographic applications}
\label{sub:qkd}

The most mature quantum communication technique currently in use is QKD.

QKD provides synchronised sources of symmetric cryptographic keys between two remote systems~\cite{qkd_wiley}. 
This technique relies on quantum physics to deliver information-theoretically secure (ITS) keys to a final cryptographic application, with the potential to replace less secure primitives once quantum computers become available.
Indeed, an ITS cryptosystem cannot be compromised even if the adversary has unlimited (quantum) computing power together with ultimately efficient algorithms~\cite{clasico_roto}.

In practice, cryptosystems based on QKD are not ITS.
Any implementation relies on assumptions that may not fully reflect the theoretical model.
In any case, all elements of the network as a whole must be as secure as possible.

Since optical quantum channels suffer from a series of limitations regarding throughput and range, a network is needed to provide end-to-end services.
A widely used approach in end-to-end key distribution is the use of so-called ``trusted nodes'', where forwarding operations can be carried out without compromising the security as a consequence of the stated trustworthiness of the locations.
In each node, a component named ``forwarding module'' (FM) performs a key transport protocol, consuming the generated key and using another ITS primitive---e.g., Vernam encryption---for ciphering and deciphering the final distributed key in each hop.

Finally, it is important to note that there are different types of QKD, particularly distinguished by their detection methods. DV-QKD, the most mature, relies on measuring discrete quantum properties of light, such as polarisation. CV-QKD, in contrast, measures quantities like phase and amplitude, and benefits from using more standardised opto-electronic components, offering potential advantages in scalability and cost.

\subsection{Production networks}
\label{sub:prod-opt-net}

When considering the realisation of quantum technologies in optical networks, it is necessary to build on a suitable model, which may differ depending on the nature of the requirements~\cite{optical_networks}.

The current networking landscape relies on several heterogeneous communication systems with different design, technology, and service fundamentals.
Besides, each of them has a different degree of interoperability.
Examples include traditional Ethernet-based local area networks; the passive optical networks (PONs) for home access; and the optical networks for aggregation and backhaul.

All operate over optical fibres and can carry the same type of traffic.
However, they have different requirements for the deployment of new systems.
In addition, network operators and service providers must comply with legal and regulatory requirements---e.g. critical infrastructure protection, lawful interception, and secrecy of communications---that prevent the deployment of novel devices unless certified and standardised.

The deployment of quantum communication would add additional requirements to these heterogeneous and already overloaded designs.

\subsection{Optical networks}
\label{sub:opt-net}

From a technical perspective, an optical network is a communication infrastructure capable of providing lightpaths.
In this context, circuits are more relevant than packets or datagrams.
Each lightpath carries optical signals that encode subscriber traffic. 

The optical network typically uses wavelength division multiplexing (WDM).
Although a bidirectional approach is possible, duplexing is usually achieved by using pairs of optical fibres, one for receiving and one for transmitting.
Transceivers encode the subscriber traffic in optical signals for specific duplexing and WDM channels.
Also, multiple elements such as filters, multiplexers---e.g., optical add-drop multiplexers (OADM)---, cross-connection systems, etc., enable the steering of these signals, while others manage their power budget ---i.e., amplifiers and variable optical attenuators.

The result is a meshed network capable of transmitting and receiving both traffic and lightpaths between nodes.
Its specific design may depend on the trade-off between performance---which includes not only throughput, but also latency, connectivity, availability---and cost.

\subsection{Optical quantum networks}
\label{sub:opt-req}

This section connects what was described so far to discuss the coexistence and integration of quantum communications in optical networks and production facilities.

\begin{figure}[h]
\centering
\includegraphics[width=0.85\columnwidth]{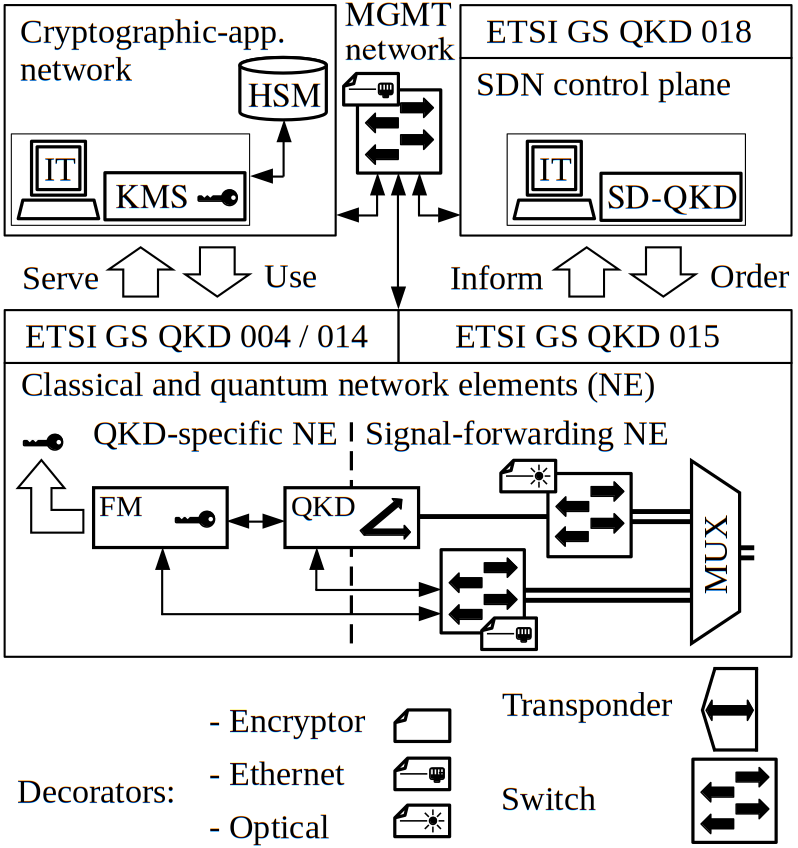}
\caption{The image shows the architecture applied to integrate quantum technology into optical networks, together with the role of ETSI GS QKD interfaces. On the left, the cryptographic network that is supported by quantum-distributed keys and has the key management system (KMS) and hardware security module (HSM) functions typical in secure scenarios. On the right, the SD-QKD control plane, which can programme the underlying quantum infrastructure. Below, the network elements (NE): i) some QKD-specific, such as the FMs that forward the key, and ii) other signal-forwarding, which enable the connectivity of the former. QKD systems can have multiple impacts, as they perform specific functions such as quantum information encoding, but they can also play an important role in optical networking techniques such as switching and power management.}
\label{fig:qnode}
\end{figure}

From the more abstract point of view, a description of a network can be obtained by outlining its functionalities, usually classified in functional planes---i.e., the forwarding, control, management, application, etc. planes~\cite{rfc_7426}.
In this regard, this work applies the concept of the ``quantum forwarding plane'' (QFP): those functionalities that enable the end-to-end quantum-enhanced provision, including FMs and supporting infrastructure.
Fig.~\ref{fig:qnode} shows the general architecture applied in the Madrid quantum network.

At present, quantum networks serve other cryptographic networks, which are based on quantum-distributed keys and integrate the key management system (KMS) and hardware security module (HSM) functions typical in secure scenarios; it is depicted on the left of Fig.~\ref{fig:qnode}. 
Interfaces ETSI~GS~QKD~004 and 014 enable the key delivery.

On the right, the SDN-type part that manages the resources of the underlying quantum infrastructure is shown, including the control and application planes typical of SDN.
The programmable interfaces, such as those defined by ETSI~GS~QKD~004, 015 or~018, enable the provisioning of end-to-end services from the network to meet the requirements of the cryptographic application network.
The implementation used in MadQCI is detailed in section~\ref{sub:oam}.

The network elements (NE) of the underlying quantum infrastructure are shown at the bottom.
These NEs include both the QKD-specific ones, such as the FMs that forwards the key hop by hop, and the signal-forwarding ones, that enable the connectivity of the former ---e.g., the shared optical components or the data network that communicates the FMs.
QKD systems can have features of both: the laser generates the pulses encoding the quantum information, but its wavelength can be changed to switch the lightpath that the signal takes through the optical network.

Finally, regarding transmission in optical networks, the quantum signals are of very low power and sensitive to extremely weak interactions~\cite{qkd_wdm_integration}.
When transmitted in a shared optical fibre, the Raman effect and four-wave mixing of adjacent channels can cause disturbances in the quantum channels.
Therefore, the allocation of the quantum and classical channels is fundamental, as is power management.
In optical networking, multiplexers and other elements are not perfect and their effects, e.g., crosstalk between channels, signal reflections, and backscattering might influence the guardband communications.
Also, active components, such as signal amplifiers, are destructive for the quantum nature of signals.
Production networks often use equipment that integrates amplification by design, such as wavelength selective switches.
As a result, these restrictions have a direct impact on both the design and technology selection for the network.

\section{The Madrid quantum-optical infrastructure}
\label{sec:qua-opt}
This section details the optical infrastructure of the Madrid network and the installation and integration of QKD systems in it, applying what has been discussed so far.

\subsection{Premises and equipment}
\label{sub:pre-equ}

What made this network unique was that it was heterogeneous and installed in real-world optical production networks, with equipment from multiple suppliers.
This section elaborates on this.

\subsubsection{Classical-specific equipment}

The infrastructure of REDIMadrid was meant for delivering IP-based connectivity services and it was outfitted with production-grade optical network devices---e.g., Ciena~6500 or Adva~FSP~3000.
RM facilities were mainly central offices or communication rooms, which fulfilled the security requirements of trusted nodes.

The premises available in the Telef\'onica Innovaci\'on Digital domain were production central offices, so they were classified as regulated critical infrastructures---e.g., taking pictures of the equipment once installed was not authorised---, suitable for validating QKD-enabled, secure communication solutions. 
This is also why the network was designed in this segment from scratch and used dedicated equipment and dark fibres. 

In both network domains, the equipment was installed in regular 19'' rack cabinets and had a power supply of 240~V, which fits the industrial standards. 
Only in the Norte and Concepción nodes---i.e., central offices of the TID domain---power inverters were used to adapt the direct current typically used in these facilities.
In all the premises, there were regular cooling conditions.

For delivering the quantum-safe services, both software and hardware resources were deployed.
The hardware devices were network encryption equipment consuming the quantum-distributed key to cipher the subscriber traffic while performing their network tasks. 
High-security Rohde~\&~Schwarz SITLine ETH encryptors were deployed for supporting ciphered L2 connectivity services. 
Analogously, some ADVA 5TCE-PCTN-10GU+AES10G modules were installed to enable L1 encryption.
Regarding the software applications, the domain-specific use cases were demonstrated using the same IT systems that ran the SD-QKD control and management system described in section~\ref{sub:oam}. 

\subsubsection{Quantum-specific equipment}

Three different QKD vendors were present in the network, namely Huawei, ID~Quantique and Toshiba. 
The ID~Quantique and Toshiba DV-QKD systems were deployed in both domains. 
The former hosted, in the same chassis, several blade-like subsystems, such as the quantum devices themselves or optical equipment for the channel adding and dropping. 
The latter ones were designed to be connected in the middle of a fibre pair that connects a facility, in the nearest point to the ducts.
It essentially bypasses all incoming or outgoing optical signal, with some losses, except for its own quantum and signalling channels. 
In terms of performance, the former achieve better performance than the latter due to the respective proprietary module characteristics.
They are not due to any specific network configuration.
The co-located set of Huawei CV-QKD systems~\cite{madrid_hwdu_switched} relayed on external optical equipment, partially shared with the ID~Quantique systems and partially standard equipment provided by the vendor.

\subsection{Quantum and classical coexistence and connectivity}
\label{sub:coe}

This section details the techniques used to enable the quantum and classical coexistence in real-world deployments. In particular, power management, judicious use of CV-QKD systems, the O-band and DWDM band allocation was used to ensure that the quantum channels remains unaffected. In addition, some of the links implemented a bidirectional scheme to dedicate a single fibre for the quantum channels.

This network connectivity is fundamental to QKD since, together with the quantum resources allocated in the optical network, the cryptographic key material has to be securely managed and transported.
Thus, there are different components that must co-operate for delivering a quantum-enabled service.
Secure design principles such as economy of mechanism ---keeping the design simple to improve security--- or access control techniques, as network slicing, helped to achieve a trusted scenario.

\begin{figure*}[!t]
\centering
\includegraphics[width=\linewidth]{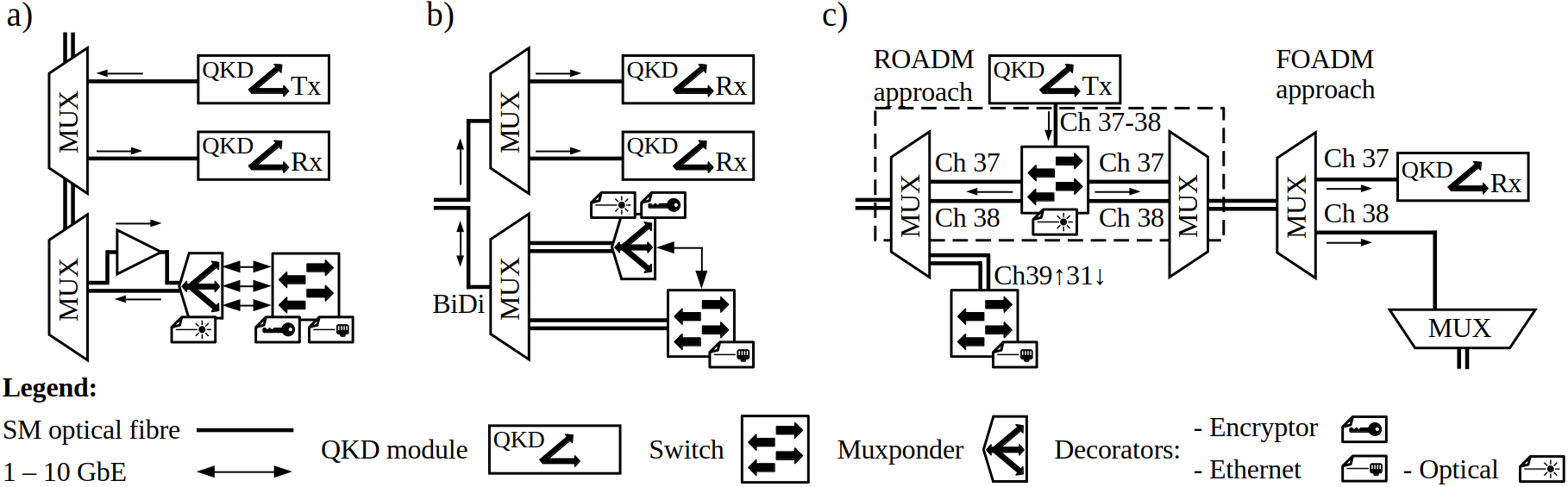}
\caption{Main coexistence schemes used in the Madrid quantum network. a) Duplex usage of a fibre pair with QKD and classical signals multiplexed in both directions. Several multiplexing stages are used to manage losses, but a single one could be used. The excess loss is compensated with an amplifier at the classical signal input. b) A bidirectional scheme is implemented with a BiDi-multiplexed fibre devoted to classical channels and the other to quantum signals. c) Example of a switched QKD network as implemented by Huawei in the co-located domain. On the left, reconfigurable-type switching is achieved by combining optical fibre switching with multiplexing (enclosed by a dashed line). Just underneath, it shows how the wide dynamic range of optical transceivers is used to pair a couple of them with different channels and better handle guardbands. On the right, fixed-type switching, where the destination is selected based on the transmission channel ---this case was not implemented and is shown for comparison purposes only.
}%
\label{fig:coexistence}
\end{figure*}

For achieving the quantum and optical integration, both quantum and classical channels were allocated in the DWDM grid using the most suitable equipment in each case as stated so far. 
However, in the network several coexistence schemes were tested, both for quantum-classical and quantum-quantum coexistence. 
Fig.~\ref{fig:coexistence} shows the main coexistence schemes implemented.

In the RM domain, the legacy production equipment could be used for aggregation and routing of signals to the terminal equipment. 
As needed, it was necessary to install ad hoc equipment for the final stages---e.g., a demultiplexer for splitting two quantum coexisting channels.
In addition, some QKD systems operated at 1310~nm instead of the 1550~nm window.
Regarding the active components, any classical signal amplification was performed after the signal aggregation, e.g., placing the amplifiers near the transceivers.

In the TID domain, the integration could be more flexible, as the equipment was not shared with other transmissions.
In the three TID nodes a higher density of equipment was deployed and coexistence was more intensively tested. 
Additionally, attack or eavesdropping emulators were installed; note that aggressive interventions such as stress test or modifying the number or power launch of classical channels would have put in risk the third-party traffic or violated the agreements to use the critical infrastructure. 
Several connectivity technologies were also tested: switch-to-switch Ethernet connectivity using coloured transceivers, both in the clear and encrypted (R\&S encryptors); OTN-like muxponders by Huawei (OTU4 graming); and the multi-frame optical encryption by ADVA modules.
On top of these, network logic was configured to ensure visibility between all QKD, IT and encrypting systems using L2/L3 network devices---e.g., ADVA FSP 150-XG304 or Huawei Quidway S6700 Series.

Additionally, a border link was implemented between both domains to enable testing cross-domain use cases. 
An ID~Quantique QKD system was deployed there and IP connectivity between the IT systems of both domains was provided in the last stage of the iteration.
Note that the co-located domain, namely the Huawei and R\&S devices, was deployed in both networks from the very beginning.

Regarding the co-located domain, Huawei implemented a switched QKD network~\cite{madrid_hwdu_switched}.
The objective was to use transmitters and receivers between any two nodes in the network as needed, allowing for a lower number of modules to be installed.
In addition, this approach makes it possible to avoid the need for trusted nodes in relatively low-loss scenarios.

\begin{figure*}[!t]
\centering
\includegraphics[width=0.9\linewidth]{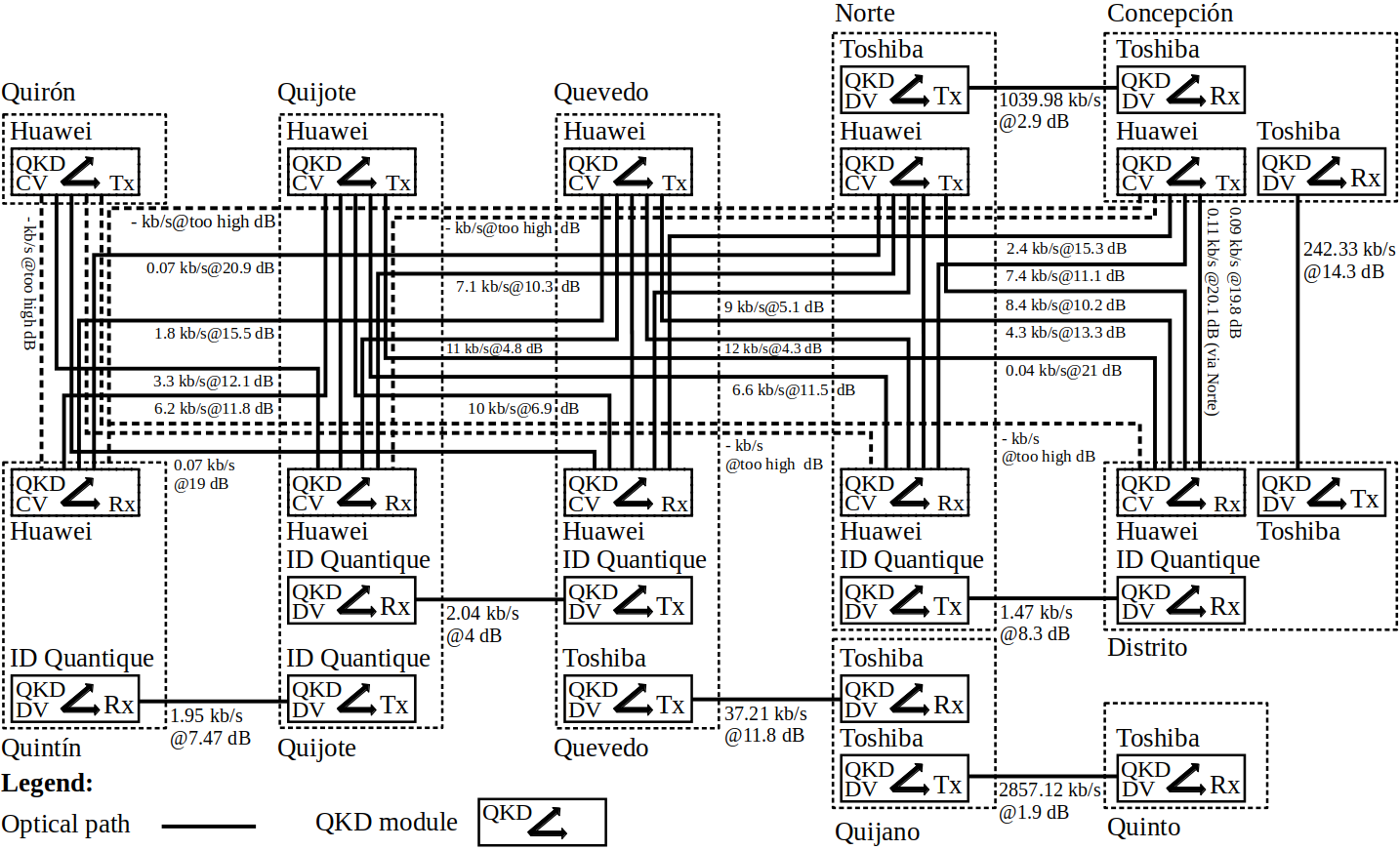}
\caption{Circuit diagram showing the possible connectivity between the QKD modules in the network. Its purpose is to visualize the two main solutions used in the network: point-to-point QKD systems and a switched QKD network. The first solution, shown at the bottom, uses DV-QKD systems from ID~Quantique and Toshiba, is less complex, and can achieve higher throughput performance ---the differences in key rates between the two manufacturers follow the respective proprietary module characteristics. The second solution, at the top, is implemented by the co-located domain with Huawei's CV-QKD systems, where switching resolves its access to the medium to achieve greater connectivity with fewer modules, except in cases with too many losses ---depicted with dashed lines. The best performing connectivity has been chosen from among the available options ---the systems can transmit on multiple channels, so different performances are obtained on the same route; see~\cite{madrid_hwdu_switched} for more details. In all cases, all circuits are implemented using a single pair of optical fibres ---and for the most part the same optical equipment---, including both quantum and classical signals. Each circuit shown, together with the generated key rate, can be exploited by the operation and management system, as explained in section~\ref{sub:oam}.
}
\label{fig:connectivity}
\end{figure*}

Each node was equipped with a fibre optic switch between the fixed optical multiplexers which, in this case, allowed channels 34, 37 and 38 to pass through.
This allows to achieve a cross-connection comparable to that of a network based on reconfigurable OADM but without the amplification typical of the wavelength selective switches.
In addition, the CV-QKD systems could tune their lasers to any DWDM channel, so any connection could be established if a sufficiently low-loss optical path was available and, at the nodes in the path, any of the three channels was not being used by another connection.
The unfeasibility of some paths, depicted with dashed lines in the Figure~\ref{fig:connectivity}, was due to the cumulative losses from the fibre distance and the successive optical stages required to aggregate and disaggregate the channels at each node. 
The achievable range may be shorter than with point-to-point solutions, but a larger number of possible connections can be achieved with fewer QKD modules: this co-located domain was able to test a total of 34 possible direct QKD link combinations between the 10 CV-QKD modules installed at 7 nodes.

In each link, optical coexistence of classical and quantum channels was achieved. 
Between 3 and 6 classical classical channels were propagated in coexistence with one DV-QKD channel and between one and four CV-QKD channels, depending on the co-located domain switching. 
Details can be found in~\cite{madrid_hwdu_switched}.
Achieving this coexistence is an outcome in itself and is therefore discussed in more detail in the section~\ref{sec:per}.

More relevant to this discussion is the connectivity achieved in the network, given that it was achieved with both point-to-point links and a switched approach.
Figure~\ref{fig:connectivity} shows that connectivity and the achieved key rate.
The point-to-point approach may perform better in terms of generated key rate, while the switched approach maximise connectivity between QKD modules. 
Thus, the network achieved both high connectivity and key rate performance, depending on the design objective.

\subsection{Operation, administration and management}
\label{sub:oam}

Several operation, administration and management (OAM) mechanisms coexisted and were tested.
Relevant to this paper is the software-defined management and operation based on ETSI~GS~QKD standards implemented at UPM, named SD-QKD Stack here.
Management systems for the classical resources were deployed in addition.

The SD-QKD Stack enabled end-to-end quantum services by configuring NEs to consume the keys generated by the links described in the previous section and forward the key material over the network.
It is composed of the SDN controller, its set of agents at the trusted nodes and a set of local key management systems (LKMS).
According to the ETSI~GS~QKD~015 specification, the last ones perform both key management and forwarding tasks.

Thus, the variety of features of the installed QKD systems in terms of key rates and standardization compliance were handled by this software, enabling their joint management and allowing the entire architecture to be exposed as a homogeneous resource driven by quality of service for applications, as specified in ETSI~GS~QKD~004~\cite{madrid_corto}.

Regarding the classical resources, in general, the optical and connectivity infrastructure was manually or automatically managed and operated using the regular techniques in each domain.

\section{Performance}
\label{sec:per}

This section details some of the most relevant results, especially with regard to coexistence and integration with optical networks in production.

\begin{figure*}[!t]
\centering
    \begin{subfigure}[t]{0.32\linewidth}
        \includegraphics[width=\columnwidth,valign=center,center]{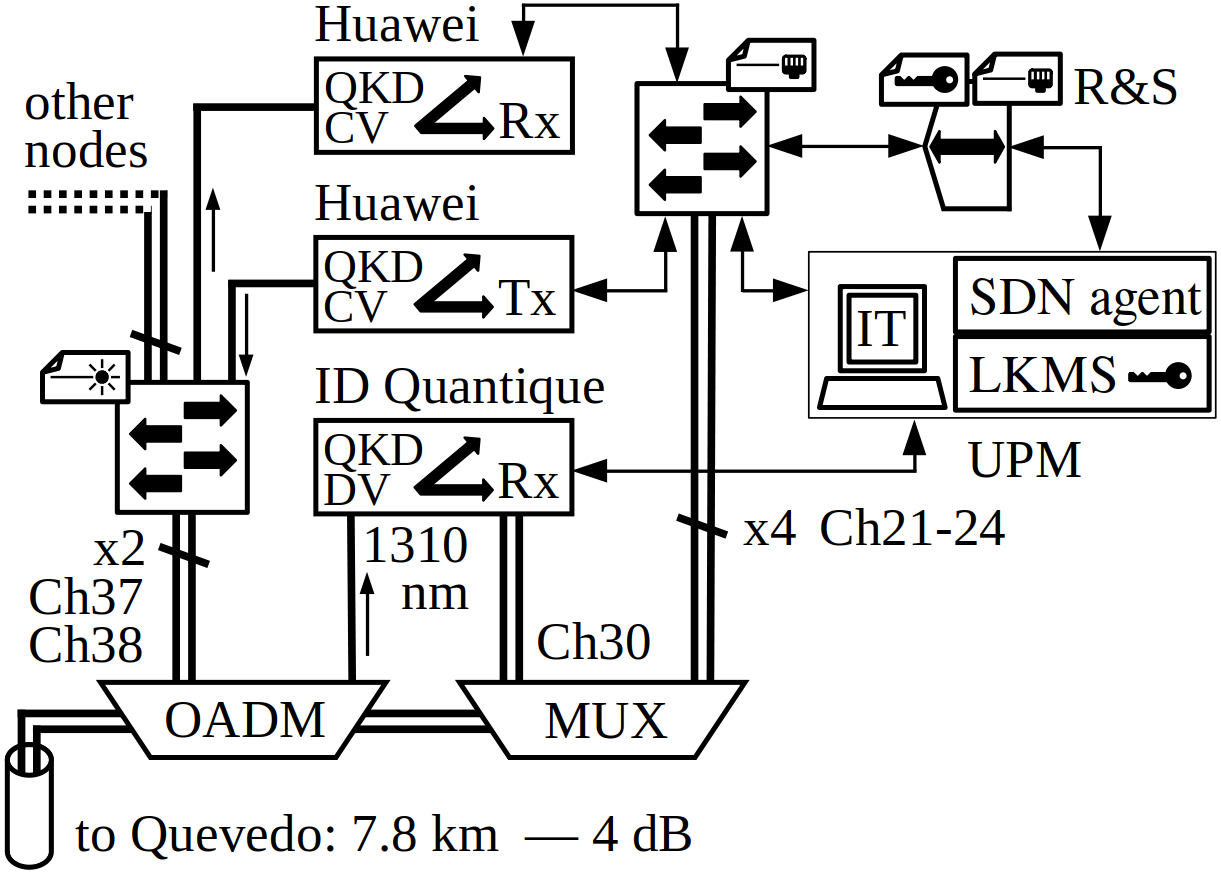}
        \caption{Coexistence scheme transmitting in the Quijote-Quevedo link. It is based on the aggregation of the quantum and classical channels by means of two multiplexers in cascade, which minimises the optical losses of the former. The legend is the same than in Fig.~\ref{fig:coexistence}.}%
        \label{fig:sch-que-qui}%
    \end{subfigure}
    \enskip
    \begin{subfigure}[t]{0.32\linewidth}
        \includegraphics[width=\columnwidth]{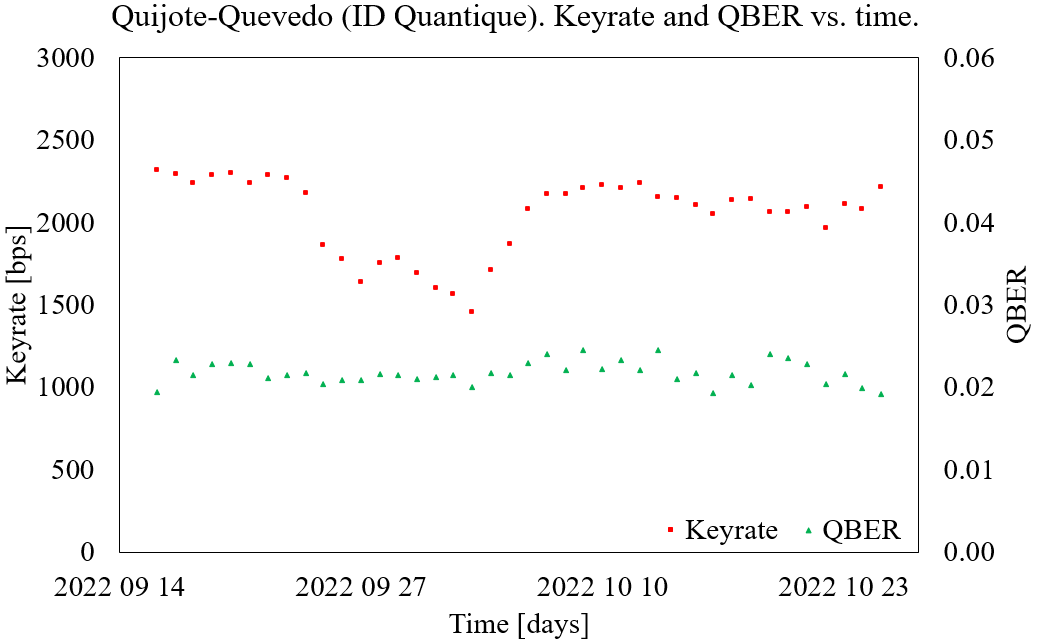}
        \caption{QBER and key rate of the ID~Quantique QKD system installed in Quijote-Quevedo. The system was installed for longer than shown, but records were only available once SNMP calls were developed to collect telemetry data.}%
        \label{fig:ber-que-qui}%
    \end{subfigure}
    \enskip
    \begin{subfigure}[t]{0.32\linewidth}
        \includegraphics[width=\columnwidth]{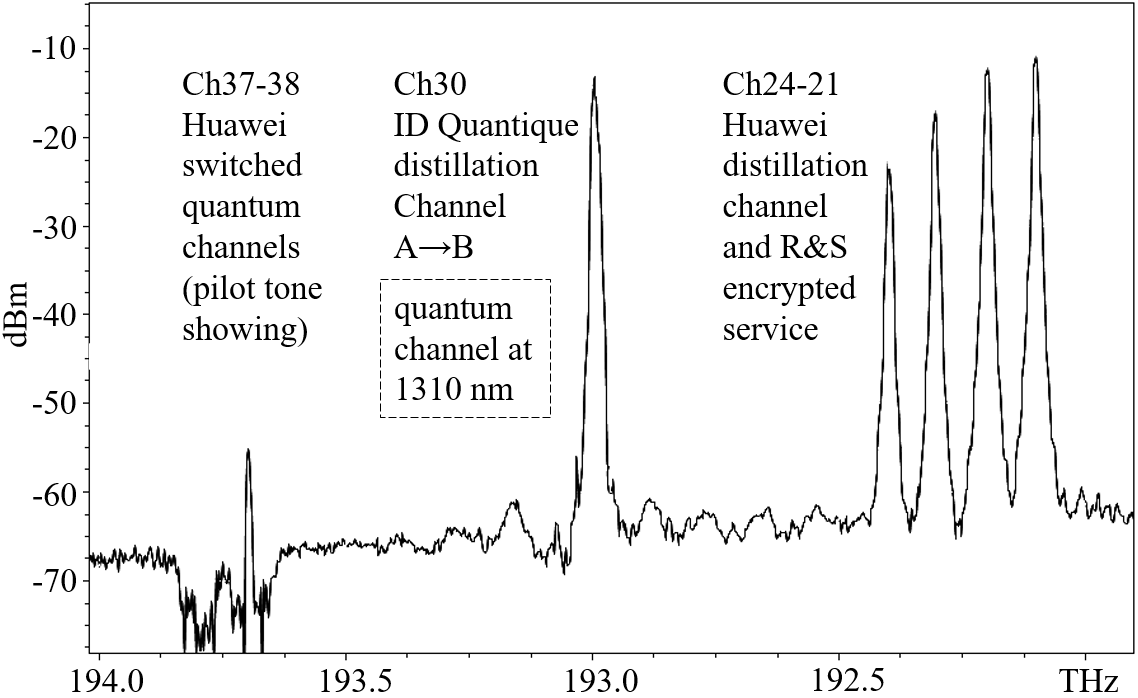}
        \caption{Spectrum of the line ingoing to Quijote from Quevedo. The notch in channels 34-38 houses the quantum signals of the CV-QKD systems, while the ID~Quantique was O-band. The outgoing line was analogous, using the same channel allocation plan.}%
        \label{fig:osa-qui-que}%
    \end{subfigure}
    \caption{Coexistence in the Quijote-Quevedo link, outfitted with a Huawei CV-QKD link and an ID~Quantique DV-QKD link.}%
    \label{fig:qui-que}%
\end{figure*}

\subsubsection{General overview}

All designs were successfully deployed and only a few minor issues related to configurations or malfunctions of different equipment had to be resolved.
Regarding the configuration, the usual complexity of any network deployment was faced, such as managing the IP addresses and the certificates for TLS-based connections.
The most usual malfunction issues were related to the information systems attached to the network and QKD equipment, both those hosting the SDN software and those that run the key management systems of each vendor ---e.g., permission and log management, software maintenance and hardware malfunctions.
There was also a complete migration of a node to another installation---Almagro to Distrito site---, precisely because it was a production central office.

\subsubsection{Quantum and classical coexistence}

Regarding the co-propagation of quantum and classical signals, two examples are highlighted in this work.
The most complex nodal schemes can be found in the supplementary material.

The first one is shown in Fig.~\ref{fig:qui-que} and represents the performance of the link between Quijote and Quevedo nodes, in which two QKD links coexisted with classical data channels, both encrypted and in the clear---only the equipment supporting this link is shown.
The Huawei continuous-variable QKD link operated in the C band---channel 37 or 38 depending on the established link--- while the ID~Quantique discrete-variable QKD link was in the O band.
The quantum-distributed key in both cases was delivered to the LKMS, to forward other key material and support the consuming applications.
In this node, an Ethernet R\&S encrypting device was installed, which is an example of these applications.

\begin{figure*}[!t]
\centering
    \begin{subfigure}[t]{0.32\linewidth}
        \includegraphics[width=\columnwidth,center]{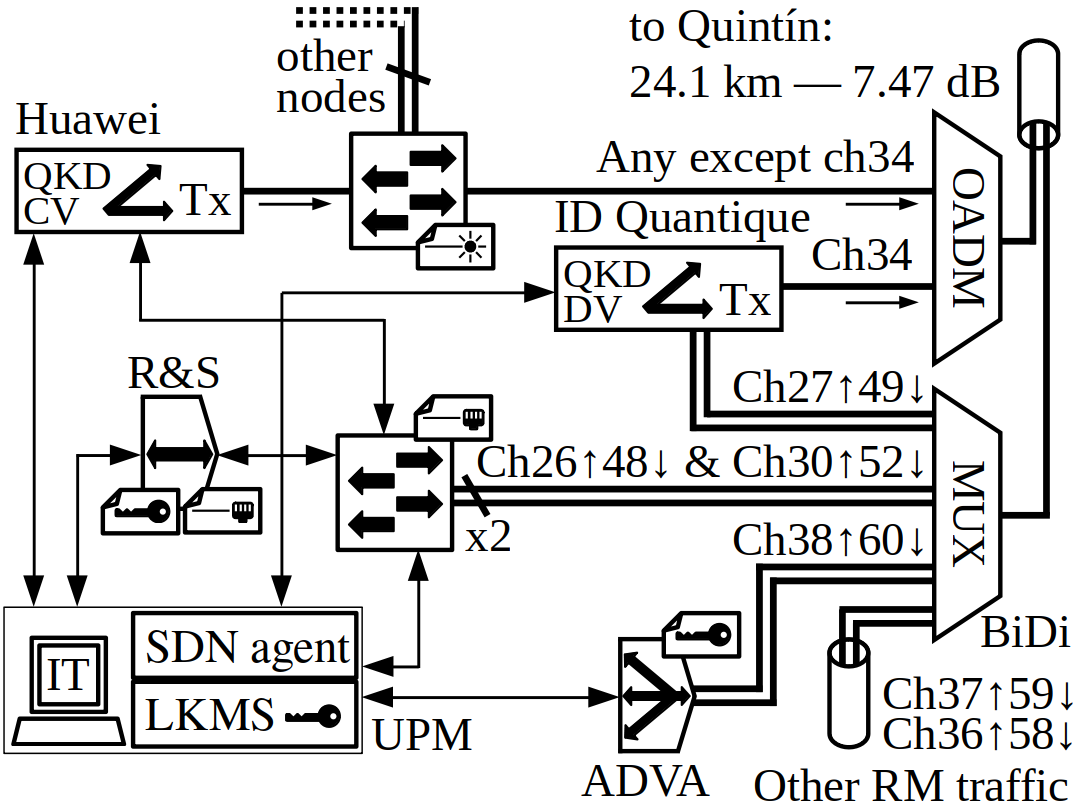}
        \caption{Coexistence scheme applied to the Quijote-Quint\'in link. It is based on using one strand for classical signals, with a bidirectional multiplexer, and the other one for quantum signals. In this case, the channels named ``other RM traffic'' includes both third-party traffic and the related to the use cases. The legend is the same as the one used in Fig.~\ref{fig:coexistence}}%
        \label{fig:sch-qui-qui}%
    \end{subfigure}
    \enskip
    \begin{subfigure}[t]{0.32\linewidth}
        \includegraphics[width=\columnwidth]{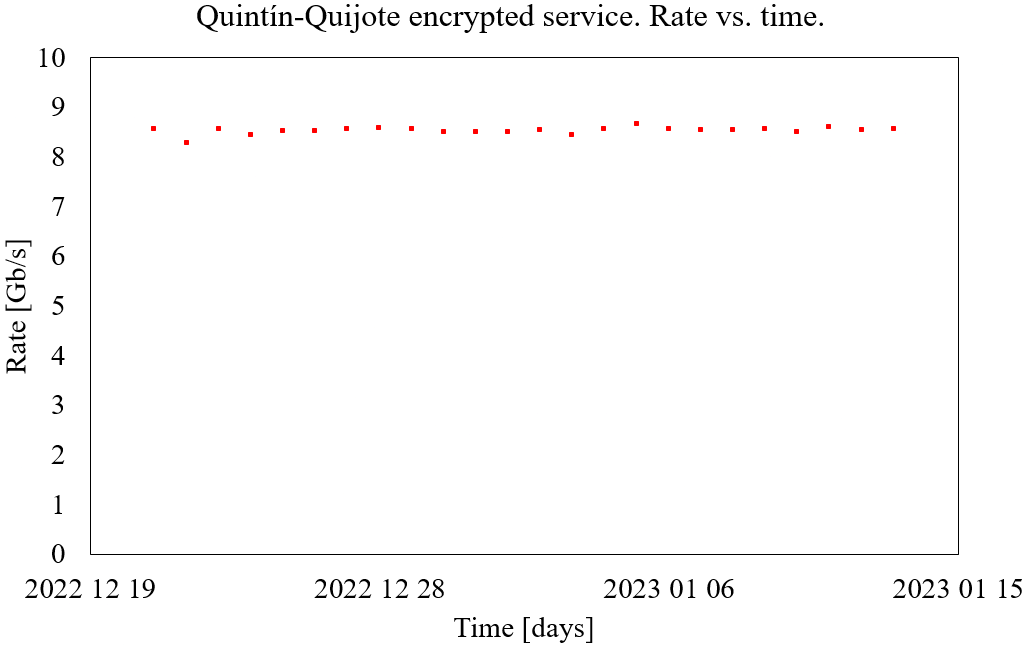}
        \caption{Performance of the L1 encryption by ADVA measured with Iperf tool. In theory, the cipher fills the frames with exactly 10~GbE/s of information for security, but this is not reflected in the host-to-host measurements due to the peculiarities of the connection, such as headers or TCP retransmissions. This ADVA encryptor is an application of the quantum network, since it consumes the quantum-distributed key to operate.}%
        \label{fig:adva-qui-qui}%
    \end{subfigure}
    \begin{subfigure}[t]{0.32\linewidth}
        \includegraphics[width=\columnwidth]{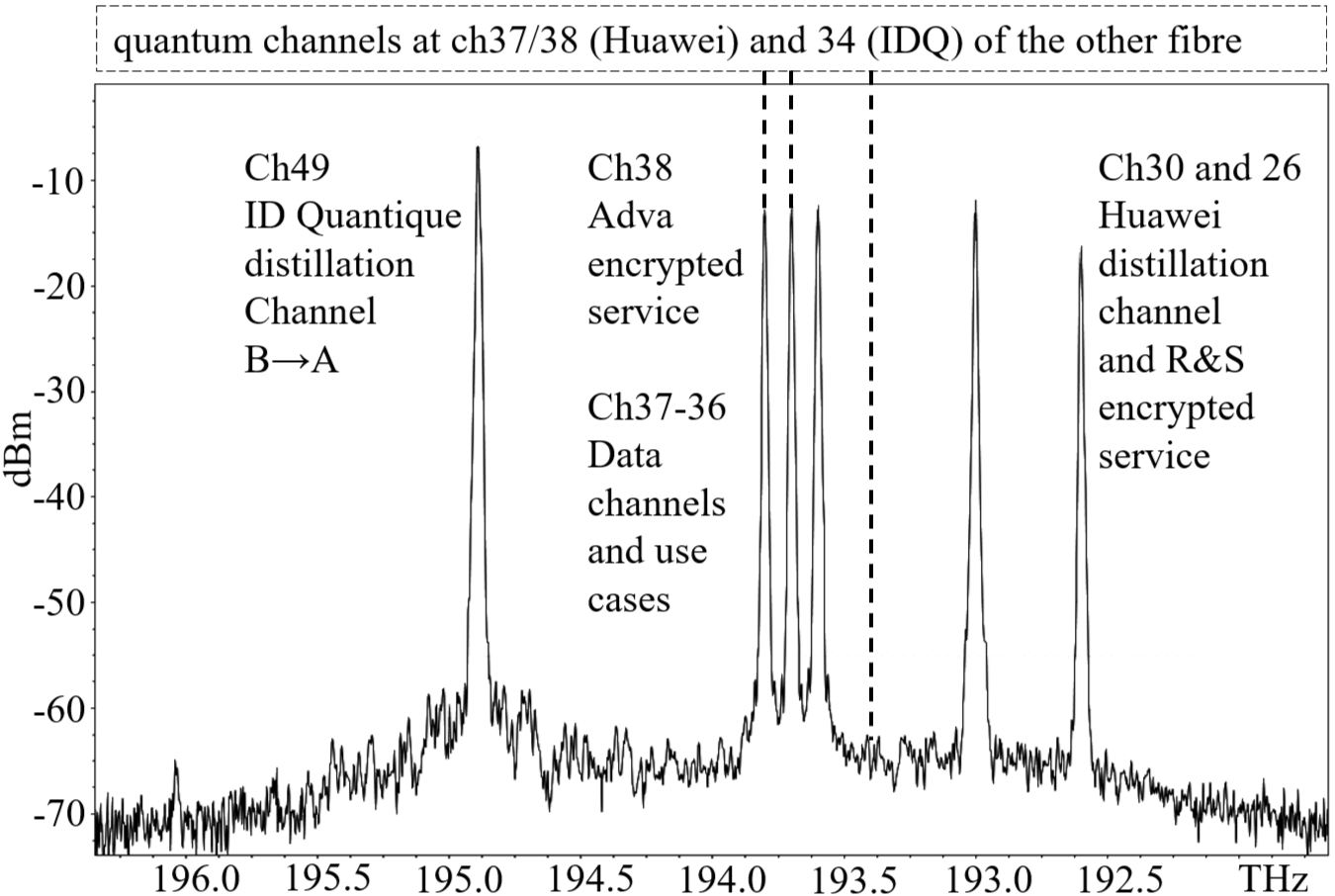}
        \caption{Spectrum of the line dedicated to the classical data channels outgoing from Quijote to Quint\'in. The channels are both encrypted (L1 by ADVA and L2 by R\&S) and in the clear. In dashed line, the position of the quantum channels at the other fibre of the pair for illustrative purposes ---in this link there is no quantum and classical coexistence in the same fibre strand.}%
        \label{fig:osa-qui-qui1}%
    \end{subfigure}
    \caption{Coexistence in the Quijote-Quint\'in link, outfitted with Huawei CV-QKD and an ID~Quantique DV-QKD links, both at C band but using a bidirectional (BiDi) approach.}%
    \label{fig:qui-qui}%
\end{figure*}

The other highlighted example includes subscribing channels.
It is depicted in Fig.~\ref{fig:qui-qui} and shows the performance of the Quint\'in-Quijote link.
Again, only the devices which transmitted or received signals from that specific link are depicted.
However, due to the switching capabilities of the co-located CV-QKD devices, quantum signals could be bypassing the node.
This link used a multiplexer for aggregating the signalling from two QKD links, Huawei and ID~Quantique, both operating in C band---channels 34 for the DV link and 37 or 38 for the CV link. 
In other links, the Toshiba discrete-variable QKD systems operated at C band while the ID~Quantique ones at both O and C bands---channel 32 the one not depicted. 
For multiplexing the classical channels, a bidirectional multiplexer was installed, and supported the signalling channels of the QKD systems, the ciphered channels of R\&S and ADVA encryptors and other RM third-party channels.

These are two examples of how the network was capable of delivering quantum-distributed keys in any link of the network while supporting third party communications in some links.
For supporting QKD, both quantum and classical channels co-propagated over the network, including distillation and synchronisation channels and data channels for the uses cases.

\subsubsection{Management, administration and operation}
The operation and management of the quantum network resources and cryptographic key was performed by the SD-QKD Stack based on ETSI~GS~QKD standards described in section~\ref{sub:oam}.
The combined operation of the controller and the sets of agents and LKMS allowed the available resources to be operated in a holistic and end-to-end approach.
This approach enabled different forwarding techniques, such as using path diversity---i.e. feeding a QKD service with several key streams---or combining keys from different vendors in each hop to provide a vendor-independent service.

Also, using the same approach, other quantum-enabled services have been trialed, such as quantum oblivious key distribution~\cite{qokd2} and quantum digital signature~\cite{qds1}.
Of interest for this work is the first case, where the raw key from some CV-QKD modules and symmetric key from DV-QKD was used to distil key material suitable for quantum oblivious transfer. 

\section{Conclusion and future work}
Achieving the deployment of heterogeneous networks in production sites is a necessary step for the widespread adoption of quantum communication. To achieve this, in addition to the known issues that limit the range of QKD systems or their coexistence with other signals, there is the added complexity of installations in real facilities that host production networks.

This paper provides a concise summary of the requirements for achieving this, as well as presenting an architecture that facilitates the integration of quantum systems into optical networks.
This approach was validated in the Madrid quantum network during the implementation of the CiViC and OpenQKD projects.

The Madrid quantum network addressed the challenge of installing 26 QKD multi-vendor modules in coexistence with classical equipment and communications, at 9 sites and 130~km of optical fibres.
In addition, it tackled several high-level issues needed for real-world quantum networking such as the joint operation and management of classical and quantum resources; the integration into the production sites of network operators and the fulfilment of their needs; the standardisation; and the quality assurance of the quantum-enhanced services provided.

Indeed, this network was deployed in a multi-domain scenario, with the support of REDIMadrid and Telef\'onica Innovaci\'on Digital. 
A multi-vendor quantum network---ID~Quantique, Huawei, Toshiba, R\&S and ADVA manufacturers---was installed with coexistence in production premises and real traffic examples.
Several quantum and classical coexistence approaches were tested, ranging from straightforward approaches to a switched QKD network.
In all cases, the premises, infrastructure elements, classical equipment and information systems were industrial and standard solutions, suitable for deployment in production sites.
Firmware updates were needed to enable the encryptors to use the quantum-distributed keys.
The performance of two of these approaches has been presented, demonstrating their feasibility for larger deployments.

The SD-QKD Stack enabled the testing of a large number of use cases.
This software implements several ETSI-QKD standards, ensuring that it is a valid solution for deployments of greater scope.

In this way, the Madrid quantum network aims to become a reference for the deployment of other QCI and EuroQCI.

\section*{Supplementary material}

Some supplementary material on this article has been gathered at a document published by Archivo Digital UPM. It includes some detailed tables and the most relevant node diagrams.

\section*{Acknowledgments}
 
The authors would like to thank projects MADQuantum-CM, funded by Comunidad de Madrid (Programa de Acciones Complementarias) and by the Recovery, Transformation and Resilience Plan—Funded by the European Union—(NextGenerationEU, PRTR-C17.I1); EuroQCI Deployment in Spain (EuroQCI-Spain), DEP grant 101091638; and the EU Horizon Europe project Quantum Secure Networks Partnership (QSNP), grant 101114043.

\bibliography{bibliography}
\bibliographystyle{IEEEtran}

\begin{IEEEbiographynophoto}{Alberto Sebasti\'an-Lombraña} is doctoral candidate at the U. Politécnica de Madrid and predoctoral assistant professor with Dept. LSIIS, E.T.S.I.Inf. He collaborates in the field of Quantum Communications Infrastructures (QCI) with the Quantum Information and Computation Group and the Center of Computational Simulation of that university. Previously, while graduating in Telecommunications Engineering (M.Sc.) at the U. Autónoma de Madrid, he collaborated with the modelling \& software engineering research group of the Computer Science Dept. of that university.
\end{IEEEbiographynophoto}

\begin{IEEEbiographynophoto}{Hans H. Brunner} is a principal research engineer at the Munich Research Center of Huawei Technologies Duesseldorf GmbH (HWDU). He is the architect and leading engineer of the CV-QKD prototypes developed at HWDU, his research focuses on CV-QKD implementations and networks. Brunner received the B.Sc., Dipl.-Ing. (M.Sc.), and Dr.-Ing. (Ph.D.) degrees in electrical engineering from Technical University of Munich (TUM) in 2005, 2007, and 2016, respectively.
\end{IEEEbiographynophoto}

\begin{IEEEbiographynophoto}{David Rinc\'on} obtained his bachelor's degree in Telecommunications from the Polytechnic University of Valladolid, Spain. Since 2002, he has worked as a Network Engineer with extensive experience in many technologies, focusing on IP and optical technologies in the service provider sector. Focused on quantum technologies, he was responsible for the deployment of the testbed in the OpenQKD project at REDIMadrid and is currently the coordinator of REDIMadrid.
\end{IEEEbiographynophoto}

\begin{IEEEbiographynophoto}{Juan~P.~Brito}
is PhD in Computer Science and currently Assistant Professor at Universidad Politécnica de Madrid. He has participated in more than 20 QKD research projects, several of them included in the FET Flagships. On these projects, his research lines have evolved around Quantum Technologies integration into modern telco infrastructures following new network paradigms like Software Defined Networking (SDN) and Network Function Virtualization (NFV). He has over 30 international publications, holds an accepted patent and is a member of the European Telecommunications Standards Institute. 
\end{IEEEbiographynophoto}

\begin{IEEEbiographynophoto}{Rub\'en~B.~M\'endez} received his Computer Engineering degree from the University of Las Palmas de Gran Canaria (UPLGC) in 2016 and then his Master's degree in 2021 from the Polytechnic University of Madrid (UPM). He is currently collaborating with the research group on quantum communications at the UPM Center of Computational Simulation where he has developed some publications in recent years. His main areas of interest are related to the insertion of technologies such as QKD in real networks through standards and tools such as SDN in order to update current protocols of classical cryptography, he also studies new ways to introduce QKD in SDN networks in a standard and transparent way for telcos.
\end{IEEEbiographynophoto}

\begin{IEEEbiographynophoto}{Rafael~J.~Vicente} received the Computer Engineer and the Computer Science Degrees from the Universidad Rey Juan Carlos (URJC) in 2018, Master's Degree in Computer Graphics, Games and Virtual Reality from the URJC in 2021 and was admitted in the Ph.D. in Software, Systems and Computing (DSSC) program from the Universidad Politecnica de Madrid (UPM) in 2021. Since 2019, he has been a Research Developer within the Group of Investigation in Quantum Information (GIICC) for the Center of Computational Simulation (CCS) of the UPM. During this period, his research has been focused on the development and implementation of standards to enable the next generation of quantum communication networks, participating in European initiatives and projects such as CIVIQ and OPENQKD. His Ph.D. thesis focuses on the study of Quantum Key Management Systems (QKMS) in Software Defined Networks (SDN) and the acceleration of QKD-assisted cryptographic algorithms.
\end{IEEEbiographynophoto}

\begin{IEEEbiographynophoto}{Jaime~S.~Buruaga} received the B.S degree in Informatic Engineering from the Complutense University of Madrid (UCM), and the M.S. degree in Networking Engineering from the Politecnical University of Madrid (UPM), in 2022. He started his PhD in 2022 in the field of Quantum Communications research group in quantum communications with the Quantum Information and Computation Group and the Center of Computational Simulation (CCS), where he is currently working as a Networking Engineer. His PhD topic is the integration of Quantum Communications in today's cryptographic networks, with several international publications. Currently, his research interest is related to the integration of standards that implement Quantum Key Distribution into current cryptographic protocols.
\end{IEEEbiographynophoto}

\begin{IEEEbiographynophoto}{Laura~Ortiz} (PhD in Physics, U. Complutense de Madrid) is currently an assistant professor at the U. Politécnica de Madrid. Her research interests covers from topological quantum Computing to quantum metrology with nitrogen vacancies. Her line of research focuses on protocols beyond quantum key distribution. She has gained experience with post-quantum cryptography algorithms recently.
\end{IEEEbiographynophoto}

\begin{IEEEbiographynophoto}{Dr~Jos\'e~Luis~Rosales} is a Senior Programme Manager in Quantum Technologies and lecturer at Universidad Polit\'ecnica de Madrid, where he leads the entrepreneurship and training line of the UPM's Quantum Communications Plan (at MADQuantum-CM). With over 25 years of experience in academia, research and innovation, he has coordinated quantum key distribution projects at national and EU levels, working with leading firms such as INDRA, Telef\'onica and GMV. He also directs the MadQ Business Venture initiative, fostering a sustainable quantum innovation ecosystem across fibre and satellite domains.
\end{IEEEbiographynophoto}

\begin{IEEEbiographynophoto}{Chi-Hang~Fred~Fung} is a principal scientist at the Munich Research Center of Huawei Technologies Duesseldorf GmbH (HWDU). He is the leading designer of QKD protocols and processes developed at HWDU. His research focuses on QKD security, QKD theory and quantum information and communications.
\end{IEEEbiographynophoto}

\begin{IEEEbiographynophoto}{Momtchil~Peev} studied mathematics and physics at the University of Sofia, Bulgaria. Ph.D. in 1993. Lise-Meitner post-doctoral Fellow at the Vienna University of Technology (1993-95) and at ARCS (resp AIT, 1995-97). Research associate at ARCS until 2010 and then senior scientist and thematic coordinator for QKD. Since 2015 he is senior expert and project leader in the Optical and Quantum Communications Laboratory at the Munich Research Center of Huawei Technologies Duesseldorf GmbH. M. Peev led the development of a QKD network prototype in the European project SECOQC (2004-2008), and coordinated the effort of the Huawei Munich Group in the SDN-QKD demonstration of Telefonica-UPM-Huawei in Madrid (2018).
\end{IEEEbiographynophoto}

\begin{IEEEbiographynophoto}{Jos\'e~M.~Rivas-Moscoso} obtained his PhD in Physics from the University of Santiago de Compostela, Spain, in 2004. From 2004 to 2010 he worked as a postdoctoral researcher at the Cambridge University Engineering Department, UK, and the School of Telecommunication of the Technical University of Madrid, Spain. In 2011 he joined the Network Core Evolution group at Telefónica I+D, Spain, focusing on beyond-100G transport technologies, and in 2014 he started his collaboration with the Network and Optical Communications group at AIT, Greece, where he investigated elastic optical networks, flexible optical switching and SDM. Since 2017, he has been with Telefónica CTIO, initially as an optical transport network architect and more recently, taking on the role of optical transport innovation expert. His research interests include multiband and SDM transmission, and QKD.
\end{IEEEbiographynophoto}

\begin{IEEEbiographynophoto}{Felipe~Jim\'enez~Arribas} holds an electronic physics degree from UCM. He joined Telefónica I+D in 1994 on a Ph.D. student scholarship, becoming part of the staff in 1995. He started working on the development of the Telefónica Spain nationwide IP network, and since then has been involved in many projects related to access and core architectures as well as in field trials of advanced fiber technologies. He has also worked on relevant EU projects, like the FP6 MUSE, FP7 Accordance, TREND, Discus, Idealist and INSPACE projects.
\end{IEEEbiographynophoto}

\begin{IEEEbiographynophoto}{Antonio~Pastor} is Industrial Engineer from the Carlos III University of Madrid (1999). He joined Telefónica I+D in 1999 working on the design and deployment of different networks worldwide. Since 2006 he has been working as an expert in network security in Telefónica. He is currently the expert responsible for security research in the network transport area in Telefonica Global CTIO Unit. He has participated in many security research programs, holding several security certifications and patents.
\end{IEEEbiographynophoto}

\begin{IEEEbiographynophoto}{Diego~R.~L\'opez} joined Telefonica in 2011 as a Senior Technology Expert, and he is currently in charge of the Technology Exploration activities within the GCTIO Unit. Before joining Telefónica he spent some years in the academic sector, dedicated to research on network services, and was appointed member of the High-Level Expert Group on Scientific Data Infrastructures by the European Commission. Diego is currently focused on applied research in network infrastructures, with a special emphasis on data-driven management, new architectures, security, and quantum communications. Diego is an ETSI Fellow and chairs the ETSI ISG ZSM.
\end{IEEEbiographynophoto}

\begin{IEEEbiographynophoto}{Jes\'us~Folgueira} is MSc in Telecommunications Engineering from UPM (1994) and MSc in Telecommunication Economics from UNED (2015). He joined Telefónica I+D in 1997. Head of Transport and IP Networks within Telefonica Global CTO unit, he leads Network Planning, Technology and Innovation. He is focused on Optical, Metro \& IP Networks, network virtualization (SDN/NFV) and advanced switching. His expertise includes Broadband Access, R\&D Management, and network deployment. 
\end{IEEEbiographynophoto}

\begin{IEEEbiographynophoto}{C\'esar S\'anchez} César Sánchez is a Research Professor at the IMDEA Software Institute. He received his PhD in Computer Science from Stanford University, USA, in 2007. He specialises in formal methods for distributed systems. After a postdoctoral fellowship at the University of California, Santa Cruz, USA, César joined the IMDEA Software Institute in 2008. He is a Scientific Researcher at the Spanish National Research Council (CSIC). Spanish National Research Council (CSIC) in 2009. In 2013, he was appointed Associate Professor at the IMDEA Software Institute, and in 2022 he became Full Professor. César holds a degree in Telecommunications Engineering (MSEE) from the Polytechnic University of Madrid (UPM), Spain, awarded in 1998. Thanks to a scholarship from La Caixa, he moved to Stanford University (USA), where he obtained a Master's degree in Computer Science in 2001, specialising in software theory and theoretical computer science. César received the 2006 ACM Frank Anger Memorial Award. His main research interests are the applications of theoretical computer science, in particular logic and games. theoretical computer science, in particular logic, games and automata theory, to the development, understanding and verification of computational devices. Professor Sánchez is also director of the REDIMadrid (RM) network. 
\end{IEEEbiographynophoto}

\begin{IEEEbiographynophoto}{Vicente Mart\'in} is a Full Professor of Computational Sciences at the Universidad Politécnica de Madrid, where he leads the Research Group on Quantum Information and coordinates the Madrid Quantum Communications Infrastructure. He has contributed to standards in Quantum Key Distribution (QKD) through organizations like ETSI and CEN, where he has been vice-chair and rapporteur in the ISG-QKD and convener in the JTC22. His research focuses on integrating Quantum Communications into telecommunications networks and security infrastructures, pioneering the use of SDN for QKD networks.
\end{IEEEbiographynophoto}

\end{document}